# SNR Optimization for Common Emitter Amplifier

Orhan Gazi
*Electrical and Electronics Engineering Department*
*Ankara Medipol University, Ankara, Turkiye*
orhan.gazi@ankaramedipol.edu.tr

*Abstract—* In this paper we investigate the effects of the thermal noise of the base resistance of common emitter amplifier (CEA) on the output SNR, and we show that a first order Butterworth filter at the output of the CEA significantly improves output SNR significantly and supress the performances of higher order Butterworth, Chebyshev I, II and elliptic filters.

We propose a formula for the selection of cut-off frequency of analog filters for given orders to achieve significant SNR improvement at CEA output. Considering the filter complexity and output SNR improvement, we can conclude that the first order Butterworth filter outperforms Chebyshev I, II and elliptic filters.

## I. Introduction

The building elements of a CEA introduce noise at different points of the amplifier. Resistors, the transistor itself, external interference and the variations in the power supplies can be considered as major noise sources. There are two major noise types available in CEA. These are shot and thermanl noises.

Shot noise is one of the noises generated at the input stage of the CBE. In a CEA amplifier constructed using a BJT, the shot noise appears at the base of the transistor due to the random fluctuations of base current. The electron flow across the depletion region of the BJT is a random process, and it is not uniform, and this causes the random fluctuations in the base current, and these random fluctuations are called shot noise. The shot noise is calculated using

$$i_{n,shot} = \sqrt{2qI_BW}$$

where we have

$q = 1.6 \times 10^{-19} C \quad electron\ charge$

$I_B\ is\ the\ DC\ base\ current$

$W\ is\ the\ bandwidth\ under\ concern$

$i_{n,shot}\ is\ the\ shot\ noise\ current$

For instance, for $I_B = 10\mu A \quad W = 10^4 Hz$ the rms shot noise at the base can be calculated using the formula

$$i_{n,shot} = \sqrt{2qI_BW}$$

as

$$i_{n,shot} = \sqrt{2 \times (1.6 \times 10^{-19}) \times (10 \times 10^{-6}) \times (10^4)}$$
$$\to i_{n,shot} \approx 5.7 \times 10^{-10} A\ (rms)$$

For an amplifier gain of 100, the shot noise at the collector equals
$$i_{n,c} = 100 \times 5.7 \times 10^{-10} \to i_{n,c} = 5.7 \times 10^{-8}\ A$$

The other major noise type is the Thermal noise generated at the resistor. Thermal noise, which is also called Johnson noise or Nyquist noise, is a type of random noise generated by the motion of electrons due to temperature in resistors and conductive materials. It occurs due to the random movement of electrons inside the resistor, which creates small voltage fluctuations.

Since amplifiers are usually designed to amplify the weak input signals, addition of noise at the base to the weak input signal can lead to significant performance degradations. The authors in [1] compare the common-base regulated cascode and common-emitter preamplifiers in terms of low-noise and wide-band performances. In [2], the authors studied switching on and off the power supply voltage of an amplifier to yield a modulated output signal that allows for synchronous demodulation, and this method improves the signal-to-noise ratio (SNR) by reducing noise. The authors in [3] analyzes the ability to suppress in-band noise power of a regenerative amplifier (RA) operating in the nonlinear regime in response to a continuous wave (CW) input tone. The authors in [4] presents an integrated low noise amplifier (LNA), based on an ac-coupled 7-stage common-emitter topology.

The noise at the base can be considered as the most important noise of an amplifier circuit, since it is amplified by the voltage gain along with the input signal, for instance, if the gain of the amplifier is 100, then the noise at the amplifier output is amplified by a factor of 100.

Larger base resistor values lead to the generation of noises with larger powers. On the other hand, the use of small base resistors may lead to early saturation problems, and larger base currents affect gain and input impedance as well. In audio, RF and biomedical applications, amplifiers are used to enhance weak input signals, and for this reason, it is critical to reduce the effects of noise signal coming from base resistors.

The thermal noise at the transistor base is due to the thermal motions of electrons in base resistors and conductive materials.

Thermal noise due to base resistors, unlike the other noise types, is always present as long as the temperature is above -273.15°C. The resistor noise, amplified noise at the output, is as a low amplitude randomly fluctuating voltage at the base of the transistor.

The voltage value of the thermal noise from the base resistor is calculated as

$$v_n = \sqrt{4kTRB} \quad (1)$$

where

$v_n$ is the RMS noise voltage
$k$ is the Boltzmann's constant, i.e., $1.38 \times 10^{-23} J/K$
$R$ is the base resistance in ohms ($\Omega$)
$B$ is the bandwidth in $Hz$

It is seen from (1) that

$$v_n \propto \sqrt{R_b}$$

and higher base resistances lead to higher thermal noise, and high values of base resistances lead to degraded output SNR. For instance, for $R_b = 100k\Omega$, $T = 27C°$, and $B = 10kHz$ (bandwidth), the RMS of base noise voltage is calculated as

$$v_n = \sqrt{4 \times (1.38 \times 10^{-23}) \times 300 \times 10^5 \times 10^4}$$

resulting in

$$v_n = 1.287\mu V$$

and for an amplifier gain of 100, the RMS of output noise voltage is

$$v_{n,o} = 1.287 \times 100 = 128.7\mu V$$

and this value is a significant value for weak signal applications like, audio and RF. In Fig. 1, we compare the effects of base resistor of the CEA on thermal and shot noise of the amplifier.

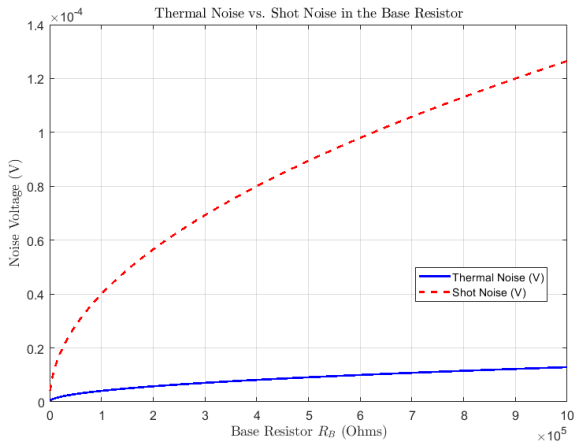

Fig. 1. Effect of Base resistor on Thermal and Shot Noise.

II. SNR INCREMENT AT THE OUTPUT OF CBE AMPLIFIER

Thermal noise is calculated as

$$P_n = Bp_n$$

where $p_n$ is the noise power spectral density.
Since thermal noise is proportional to bandwidth, it is possible to decrease the noise power at the output by reducing the bandwidth, and this can be achieved using low-pass filters. The low-pass filters concatenated to the output of the amplifier increases output SNT but it cannot fully eliminate the added noise.

*A. CBE Amplifier with Low-pass Filters*

In Fig. 2, we see the SNR improvement when Butterworth low-pass filter is concatenated to the CBE amplifier. It is seen from Fig. 2 that using large order Butterworth filters for a given cutoff frequency has negative effect on SNR improvement. For a input signal frequency of 50Hz, considering the hardware complexity of the circuit, it is seen from Fig. 1 that a first order Butterworth filter with cutoff frequency of 600Hz is the preferable choice, since it has lower hardware complexity and lower cost.

In Fig. 3, the SNR improvement at CBE amplifier output for a Butterworth filter of order 4 for different cutoff frequencies is depicted. It is seen from Fig. 3 that for higher order Butterworth filters larger cutoff frequency is needed to achieve significant SNR improvement.

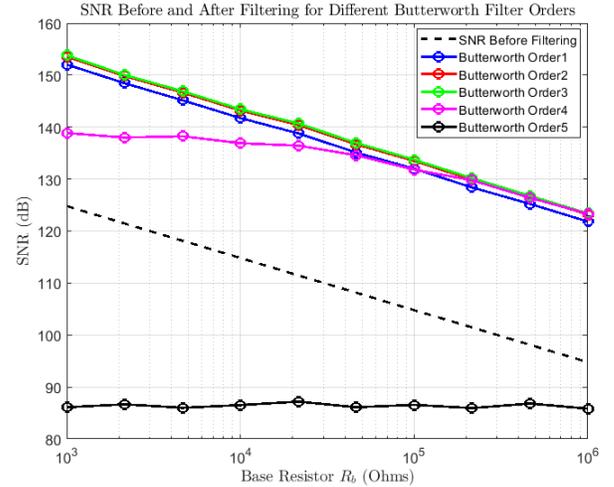

Fig. 2. CBE Amplifier Output SNR before and After Filtering, Source signal frequency is 50Hz, Butterworth cutoff frequency is: 600Hz.

It is seen from Fig. 4 that if very large cutoff frequency is used for all the Butterworth filters which have different orders, they show very similar improvements on output SNR. However, if the low hardware complexity and cost is of the primary concern, then it is wise to choose the lowest order, i.e., first order, Butterworth filter for SNR improvement.

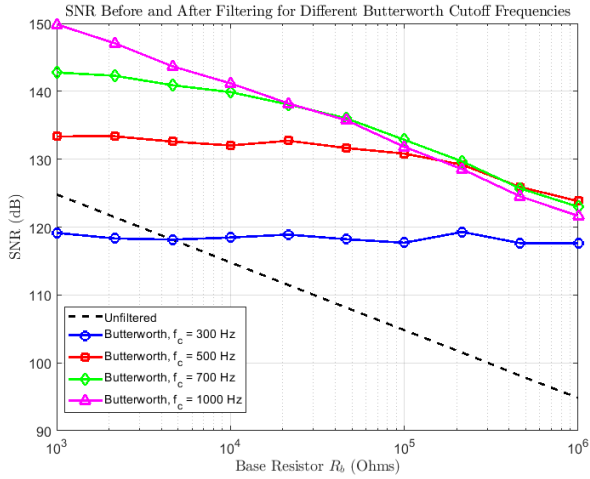

Fig. 3. CBE Amplifier Output SNR after Butterworth Filtering with Different Cutoff Frequencies, Source signal frequency is 50Hz, Filter Order N=4.

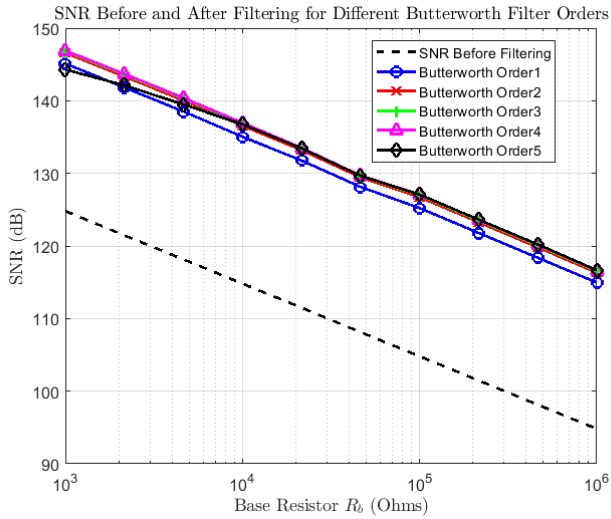

Fig. 4. CBE Amplifier Output SNR before and After Filtering, Source signal frequency is 50Hz, Butterworth cutoff frequency is: 3 kHz.

Comparing the Figures-2, 3, and 4 we can conclude that for source frequencies less than 1kHz, the cutoff frequency of the Butterworth filter for significant SNR improvement can be chosen as

$$Butterworth\ cutoff\ frequency \approx 2^{Filter\ Order+1} \times Signal\ Frequency$$

and for source frequencies more than 1kHz the cutoff frequency of the Butterworth filter can be chosen as

$$Butterworth\ cutoff\ frequency \approx 2^{Filter\ Order-2} \times Signal\ Frequency$$

In Fig. 5, the SNR improvement at CBE amplifier output for different ordered Butterworth filters with cutoff frequency of 3kHz and for an input frequency of 1kHz is depicted. It is seen from Fig. 5 that all the filters show similar performance.

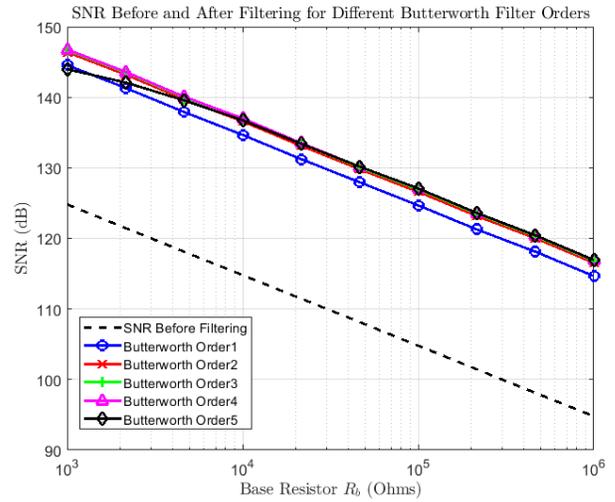

Fig. 5. CBE Amplifier Output SNR before and After Filtering, Source signal frequency is 1 kHz, Butterworth cutoff frequency is: 3 kHz.

In Fig. 6, we consider the effects of Chebyshev Type I filter on the SNR improvement at CBE amplifier output for different cutoff frequencies are displayed. As the cutoff frequency increases, better SNR improvement is observed, and this is the expected result. Chebyshev Type II and Elliptic filters show similar performance as depicted in Fig. 7 and Fig. 8.

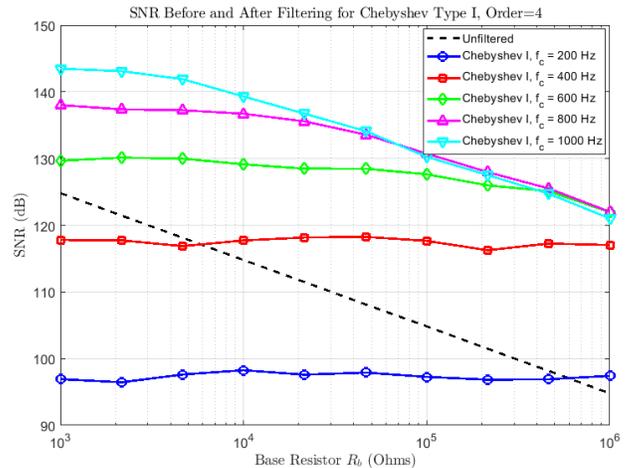

Fig. 6. CBE Amplifier Output SNR before and After Filtering, Source signal frequency is 50 Hz, Chebyshev Type I filter order =4, Passband Ripple Rp = 0.5 dB.

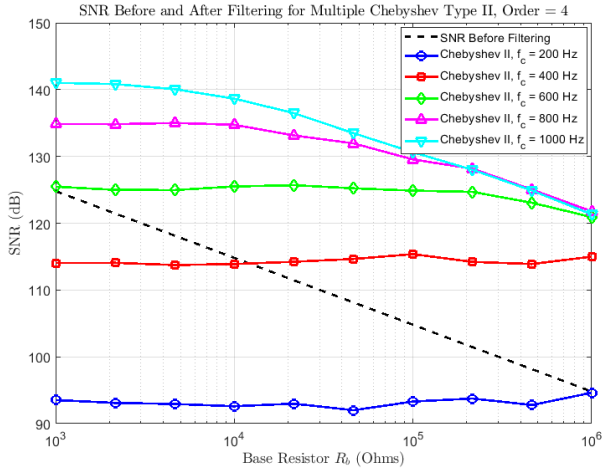

Fig. 7. CBE Amplifier Output SNR before and After Filtering, Source signal frequency is 50 Hz, Chebyshev Type II filter order =4, Stopband Attenuation Rs = 30 dB.

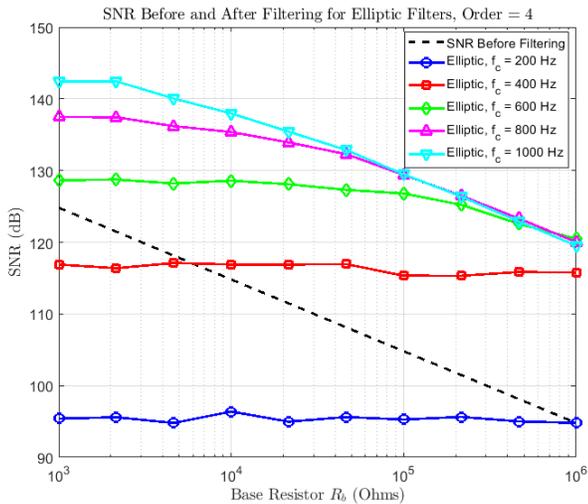

Fig. 8. CBE Amplifier Output SNR before and After Filtering, Source signal frequency is 50 Hz, Elliptic filter order =4, Passband Ripple Rp = 0.5dB, Stopband Attenuation Rs = 30dB

In Fig. 9, we compare the performance of first order three types of filters, and it is seen from Fig. 8 that Butterworth filter is the best choice to be employed at the output of the CBE amplifier to improve SNR.

### III. CONCLUSION:

Thermal noise at the base of the transistors due to the base resistors is a critical issue, since it appears as an amplified noise at the output of the amplifier. In this paper, we considered the integration of low-pass filters to the output of the CBE amplifier to improve the output SNR.

The common belief in literature is that higher order filters always give better performance, however, they suffer from high complexity implementations. Contrary to the common beliefs, it is seen from the simulation results that employment of the first order Butterworth filter is the most convenient approach to improve the output SNR considering the complexity performance tradeoff with respect to Chebyshev and Elliptic Filters, and the implementation of such filter both in hardware and software has the lowest complexity.

We also proposed a formula for the selection of cutoff frequency of a filter for a given order and source frequency to improve the output SNR

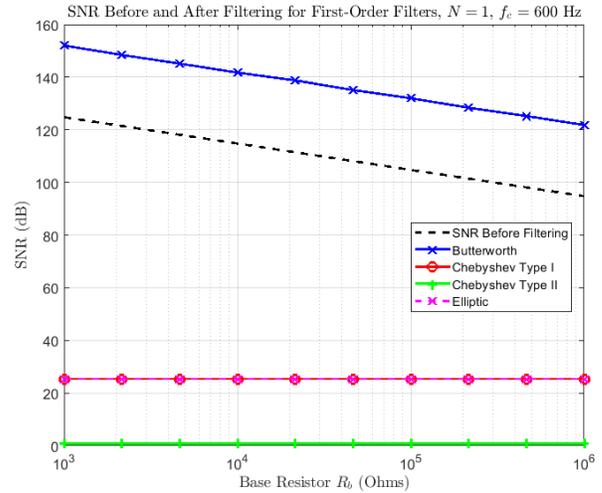

Fig. 9. CBE Amplifier Output SNR before and After Filtering Source signal frequency is 50 Hz, Order=1 Passband Ripple for Chebyshev I & Elliptic Filters: Rp = 0.5dB Stopband Attenuation for Chebyshev II & Elliptic Filters: Rs = 30dB

REFERENCES

[1] C.B. Yahya, "Design of wideband low noise transimpedance amplifiers for optical communications," *Proceedings of the 43rd IEEE Midwest Symposium on Circuits and Systems*, 08-11 August 2000, Lansing, MI, USA)

[2] G. Hornero, O. Casas and R. Pallàs-Areny, "Signal to noise ratio improvement by power supply voltage switching," *16th IMEKO TC4 Symposium, Exploring New Frontiers of Instrumentation and Methods for Electrical and Electronic Measurements*, Sept. 22-24, 2008, Florence, Italy

[3] B. H. Lam, P. Gudem, P. P. Mercier, "Analysis and Measurement of Noise Suppression in a Nonlinear Regenerative Amplifier," *IEEE Transactions on Circuits and Systems I,* volume: 69, issue: 10, october 2022

[4] P. Stärke, L. Steinweg, C. Carta; F. Ellinger "Common Emitter Low Noise Amplifier with 19 dB Gain for 140 GHz to 220 GHz in 130 nm SiGe," *2019 International Conference on Wireless and Mobile Computing, Networking and Communications (WiMob)*, 21-23 October 2019, Barcelona, SpainK. Elissa, "Title of paper if known," unpublished.